# Miniaturized Chaos-assisted Spectrometer

Yujia Zhang[1,†], Chaojun Xu[1,†], Zhenyu Zhao[1], Yikai Su[1,*], Xuhan Guo[1,*]

[1] State Key Laboratory of Photonics and Communications, School of Information and Electronic Engineering, Shanghai Jiao Tong University, Shanghai 200240, China

*Corresponding author. Email: guoxuhan@sjtu.edu.cn; yikaisu@sjtu.edu.cn.

†These authors contributed equally to this work.

**Abstract**

**Computational spectrometers are at the forefront of spectroscopy, promising portable, on-chip, or in-situ spectrum analysis through the integration of advanced computational techniques into optical systems. However, existing computational spectrometer systems have yet to fully exploit optical properties due to imperfect spectral responses, resulting in increased system complexity and compromised performance in resolution, bandwidth, and footprint. In this study, we introduce optical chaos into spectrum manipulation via cavity deformation, leveraging high spatial and spectral complexities to address this challenge. By utilizing a single chaotic cavity, we achieve high diversity in spectra, facilitating channel decorrelation of 10 pm and ensuring optimal reconstruction over 100 nm within an ultra-compact footprint of 20×22 μm$^2$ as well as an ultra-low power consumption of 16.5 mW. Our approach not only enables state-of-the-art on-chip spectrometer performance in resolution-bandwidth-footprint metric, but also has the potential to revolutionize the entire computational spectrometer ecosystem.**

## Introduction

Chaotic systems, whose eventual behavior is exponentially sensitive to the small deviation of initial condition, is a fundamental phenomenon manifesting random behaviors across various fields including astronomy[1], chemistry[2], biology[3], etc. In optical systems, chaotic behavior can lead to the unpredictable photonic phenomenon with high spatial and spectral complexities, where numerous optical motions overlap in spectrum and spatial distributions within the chaotic photonic system[4]. The intrinsic properties of chaos raise a full-new potential for steering the system to the desired final state at minimal cost, dubbed the control of chaos[5–8]. A classic example of chaotic photonics systems is the model system for wave chaos on dielectric microcavities in which the deformed microcavity can be conceptualized as a two-dimensional billiard with total reflecting walls where the light wavelength is much smaller than the microcavity scale. The smooth boundary deformed approach has been considered for chaos analysis in phase space and ray dynamics. It is implemented to describe the chaotic motion of waves where a small deviation would result in the exponential divergence of trajectories[9,10]. Different from typical circular microcavities where only whispering-gallery modes (WGMs) are supported due to the phase matching, these chaotic microcavities can provide more complex resonant modes[10–13]. In the past decades, the direct observation of chaotic motions[11,14], exploration of fundamental physical phenomena in chaotic cavity[12,13,15–17], and chaos-assisted applications[18–20] have been demonstrated. These endeavors often utilize effective chaos control methods to attain desired results at minimal expense. What fascinates us is the direct utilization of chaotic phenomena for generating random and chaotic information, addressing specific application requirements due to delivering a high level of resilience and confidentiality, such as transmission carriers in optical communications[21], random bit generators[22,23] and image encryption[24]. In contrast to chaos control, which explores novel physical phenomena within chaotic systems, the intrinsic behavior of chaotic systems is directly harnessed for these specialized fields requiring randomness and unpredictability in the targets. Computational spectrometer stands out as a prime example due to its desperation for a random response matrix, yet the idea of employing chaotic spectra in spectrometry had never been explored, offering a brand-new paradigm of optical spectrometer miniaturization requirements for portable, on-chip, or *in-situ* spectrum analysis.

Optical spectrometers is a crucial characterization technique utilized across various disciplines, wherever light can effectively probe matter, serving a pivotal role for analysis and investigation in the biological, physical, and astronomical sciences[25]. Leveraging embedded computing, computational spectrometers transcend traditional spectrometers' performance constraints by utilizing algorithms to deconvolute light interactions with spectral encoders, thereby reconstructing the original spectrum. This process relies on the random matrix, which is fundamental to compressive sensing (CS) and underpins the acquisition and processing of spectral data in computational spectrometers. Randomness, or high degree of decorrelation, contributes to a well-conditioned measurement matrix with a low condition number, critical for reducing reliance on prior knowledge as well as validating the stability and accuracy of spectrum reconstruction. Unlike conventional optical systems that aim to eliminate disorder and chaotic behaviors of light transmission, computational spectrometers introduce randomness and diversity through complex spectral-to-spatial mapping[26–28] or spectral response engineering[29–32]. Yet despite these attempts, none have managed to overcome the three-way trade-off between resolution, bandwidth, and physical size due to imperfect spectral responses. Achieving high spectral resolution often comes at the expense of limited bandwidth and necessitates a long optical path length, leading to a larger footprint required for sufficient spectral decorrelation. Recently Yao *et al.*[33,34] achieved ultra-high bandwidth-to-resolution ratios but require complex cascading configurations and up to millimeter-scale footprints for uncorrelated sampling, exposing constraints on miniaturization and integration of spectrometers. Furthermore, particularly for on-chip spectrometers based on resonators, it is the periodicity of the response matrix limits operational bandwidth or spoils orthogonality, resulting in large condition numbers. One strategy involves incorporating few transmission modes with slight offsets, realized through higher-order mode[35], dispersive coupling[36], or circular microdisk[37]. Yet, this approach only partially mitigates the periodicity issue, resulting in more stringent stability requirements for test environment and sharply increased demands for computing resources. To this end, chaotic photonics systems allow for the generation of chaotic spatial or spectrum distributions with high periodicity suppression in a single microcavity, offering an encouraging medium for ultra-compact footprint and high integrability.

In this work, we present a novel chaotic computational spectrometer that capitalizes on intrinsic chaotic behavior. Unlike conventional methods, our approach utilizes a single chaotic cavity to fully utilize the chaotic spectral information. The design of the chaotic cavity involves deforming its boundary into a Limaçon of Pascal shape, exploiting chaotic behavior to effectively eliminate periodicity in resonant cavities. With proper degree of deformation, the chaotic system exhibits a mixed phase space where chaotic and regular regions coexist, in which more resonant modes emerge to complicate the spectrum distribution. Experimentally, we demonstrate a chaos-assisted computational spectrometer achieving ultra-high resolution (10 pm) and broad operational bandwidth (100 nm) in a single chaotic cavity, facilitated by a highly de-correlated response matrix with effectively suppressing periodicity. Additionally, the footprint of the spectrometer is compacted to a mere 20×22 μm², in the meantime addressing the three-way trade-off of resolution-bandwidth-footprint metric in prior-art spectrometers, and rendering it highly suitable for miniaturized *in-situ* sensing systems.

**Results**

**Principle**

The chaotic spectrum is based on an individual asymmetric microdisk resonator which undergoes a smooth deformation of the boundary. The boundary curve of the cavity is described by the Limaçon of Pascal which is expressed in polar coordinates as $\rho(\varphi) = R(1 + \alpha\cos\varphi)$, where the deformation parameter $\alpha = 0.375$ with an effective radius $R$ of 10 μm. As the deformation parameter $\alpha$ approaches zero, the shape asymptotically resembles a circle with a radius $R$. Fig.**1a** and **1b** depict the schematics of the circular microdisk cavity and the chaotic cavity, respectively. The chaotic motion of photons in our silicon (Si) cavity is analyzed through exploiting the Ray model[38–40], where the size of the proposed chaotic cavity surpasses the limit imposed by the short wavelength of the wave. Ray dynamics is addressed in a cavity with a perfectly reflecting boundary to concisely reveal the real-space trajectories, and the Poincaré surface of section (PSOS) of this chaotic cavity is derived by tracing the light as geodesic lines reflecting at the cavity boundary. Fig. **1c** and **1d** illustrate PSOSs of the circular microdisk and of the chaotic cavity, respectively, with Birkhoff coordinates $(S, \sin\chi)$, where $S$ is the arclength of the cavity boundary, $\chi$ is the incident angle and $S_{max}$ is the total round of the cavity boundary. The calculated ray trajectories in real space, which are trapped by total internal reflection (TIR), are marked with different colors in the PSOS and are plotted in Fig. **1c** and **1d**. The condition for TIR is delivered by $\sin\chi > 1/n$, where *n* denotes refractive index, and the black dashed line in PSOS marks the boundary of the TIR and leaky region. For the circular microdisk, rotational invariance of the system ensures conservation of the angular momentum. Stable periodic orbits completely fill the PSOS of the circular microdisk. Due to the inability to enter the leaky region, a ray remains confined within the cavity indefinitely. As presented in bottom part of Fig. **1c**, the circular microdisk can only support a limited number of orders of WGMs. In contrast, for the chaotic cavity, the chaotic trajectories are cut into infinitely dispersed dispersion on the PSOS, which is called chaotic sea. As scattered dots marked in red in the PSOS in Fig. **1d**, the corresponding ray trajectory in real space is provided in the upper row with red lines in chaotic and disordered configuration. Periodic and quasi-periodic trajectories are cut into finite points or closed curves embedded in chaotic sea, which are called islands. Blue (periodic mode) and green (quasi-periodic mode) lines provide the corresponding ray trajectories in real space, as indicated by the blue and green dots. Due to the complex mixed phase space structure, the chaotic cavity naturally supports a wide variety of resonance modes, as exhibited in bottom part of Fig. **1d**, encompassing resonances confined by stable periodic orbits owing to TIR, as well as wave localization within the chaotic sea, coexisting simultaneously. As the angular momentum of light in the chaotic channel approaches that of stable orbital modes, efficient coupling can be achieved via dynamic tunneling[13]. This process operates independently of the phase-matching condition compared to the circular microdisk cavity restricted to only support WGMs[10–12], allowing a larger number of resonant modes in the chaotic cavity. The presence of intricate resonant modes in the chaotic cavity suggests a complicated spectrum response with a high degree of diversity and indicates weakened periodicity compared to the circular microdisk cavity which exclusively supports WGMs in a few specific orders. The dynamics of the momentum transformation are further explored. The chaotic motion converts the angular momentum of light into stable orbital resonant mode pattern within a few dozen picoseconds, demonstrating excellent temporal stability for spectral matrix reconstruction, as detailed in Supplementary Information **S1**.

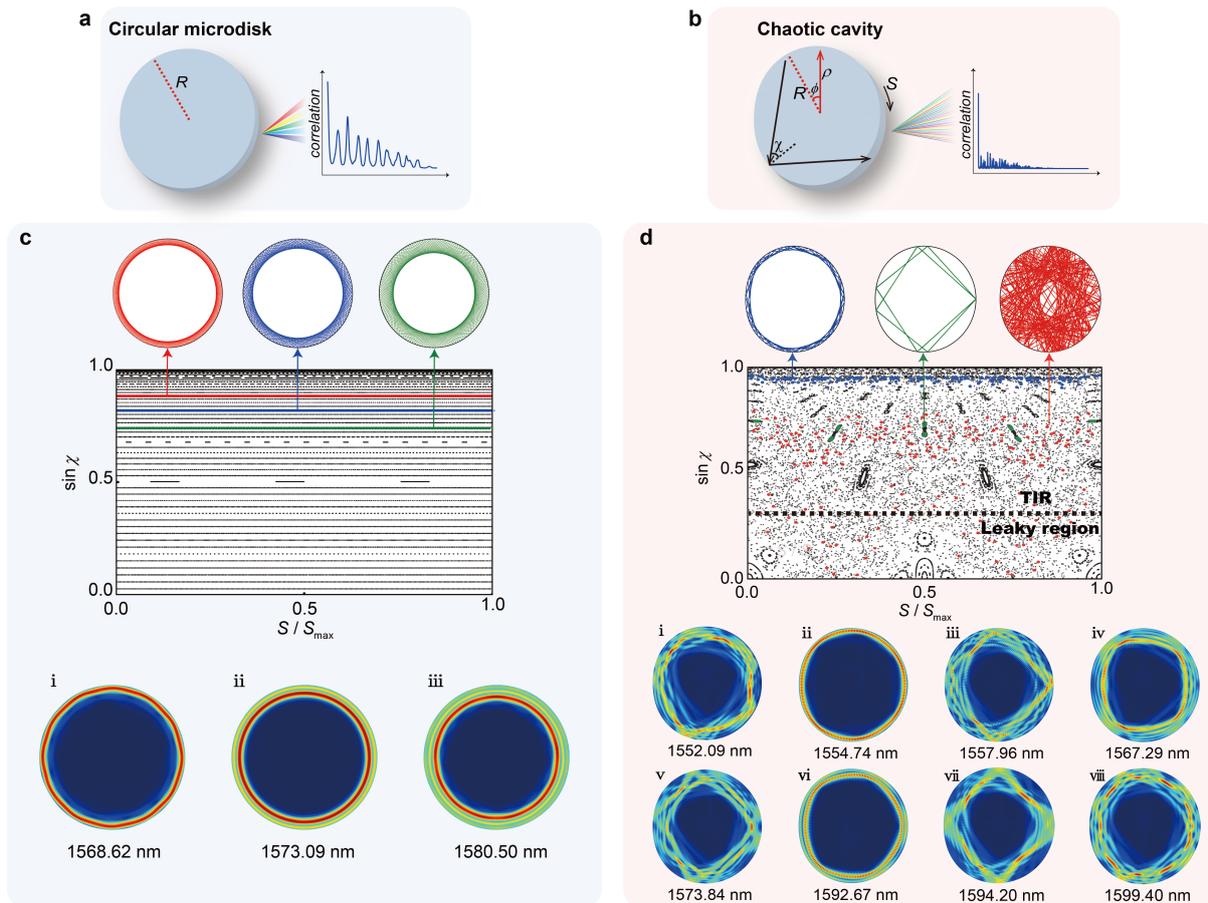

**Fig. 1 | Chaotic cavity and circular microdisk cavity. a,** schematic of circular microdisk and its strong periodicity visualized by correlation function, **b,** schematic of chaotic cavity and its weak periodicity visualized by correlation function. The calculated PSOS, the corresponding ray trajectory in real space marked with different colors in PSOS, and supporting resonant modes of **c,** the circular microdisk cavity and **d,** chaotic cavity, respectively.

**Chaotic spectra analysis**

The chaotic effects of periodicity suppression and diversity enhancement of spectra induced by deformation parameter $\alpha$ are analyzed. We first calculate the PSOS of these chaotic cavities with different $\alpha$ ($0.3 \leq \alpha \leq 0.5$). Fig. **2a** depicts the boundary and corresponding PSOS. (Refer to Supplementary Information **S2** for the initial scanning of deformation parameter over a coarse range, with $\alpha$ ranging from 0 to 0.5). It's observed from the classical ray dynamics that increasing the deformation parameter results in the disappearance of more invariant curves and to an increase of chaotic performance. The region of the chaotic sea expands, and the islands of stability corresponding to periodic/quasi-periodic orbits gradually diminish until they vanish completely, fully occupied by the chaotic sea when $\alpha \geq 0.4$. The spectral responses of these fabricated chaotic cavities with an add-drop configuration are characterized from drop ports (refer to Materials and methods for measurement details). The auto-correlation function of spectral response is used to quantitatively evaluate the periodicity of these chaotic cavities. We employ the maximum value following a descent from the initial value of '1' in the normalized auto-correlation function as the figure-of-merit (F.O.M) for periodicity quantification. The F.O.M reveals the degree of column orthogonality (independence of wavelength channel) in the response matrix. A larger F.O.M suggests that redundant information is present in the response matrix, leading to increased multicollinearity, which can induce instability in the solution and result in poor reconstruction. The periodicity as a function of the deformation parameter $\alpha$ is shown in Fig. **2b** as red points (refer to Supplementary Information **S4** for measured transmission spectra and calculated auto-correlation functions of each device). As $\alpha$ increases to 0.375, the spatial symmetry of the chaotic cavity is more thoroughly broken, and chaotic effects are strengthened. Chaotic modes and various periodic/quasi-periodic resonant modes are delivered simultaneously; thus, diversity and disorder are effectively introduced into the spectral responses. In this way, periodicity can be significantly suppressed to around 0.23 as $\alpha = 0.375$ compared to 0.55 for circular microdisk cavities (refer to Supplementary Information **S3** for the spectral response, and auto-correlation function for the circular microdisk cavity). The insertion loss of these chaotic cavities is evaluated as well and exhibited in Fig. **2b** as blue dots. When $\alpha \leq 0.375$, the insertion losses increase slowly and modestly due to the existence of stable islands. However, as $\alpha$ exceeds 0.4, the islands of stability shrink to vanish, leading to the cessation of the corresponding resonant modes. These stable trajectories of original resonant modes convert to chaotic trajectories which traverse the leaky region, resulting in escape events and high loss. Lazutkin's theorem states that a convex and smooth billiard boundary always has invariant curves at $|\sin\chi| \approx 1$[41]. Hence, a small number of resonant modes with high $\sin\chi$ persist. Consequently, these limited modes not only maintain considerable losses but also conversely magnify the periodicity. Considering both the periodicity suppression effect and insertion loss, $\alpha = 0.375$ is selected as the optimal deformation parameter for our chaos-assisted spectrometer. Furthermore, we explore how the gaps between the chaotic resonator and bus waveguides within coupling regions affect the Q-factor, thereby influencing the resolution of the spectrometer. The Q-factor dictates the wavelength-channel decorrelation through thermo-optical tuning in silicon photonics, with a higher Q-factor enhancing spectral resolution[36]. The loaded Q-factor of a resonator is co-determined by two components as $Q_{load}^{-1} = Q_{int}^{-1} + Q_{ext}^{-1}$, where $Q_{int}$ depends only on the intrinsic cavity loss and $Q_{ext}$ signifies the Q-factor reduction due to the external power loss caused by evanescent coupling to the bus waveguides. We fabricated a series of chaotic cavities with the same $\alpha$, identical effective radius of 10 μm and varying coupling gaps from 80 nm to 180 nm. The estimated resolution can be obtained by the spectral correlation function of spectral response:

$$C(\Delta\lambda, N) = \frac{\langle T(\lambda, N)T(\lambda + \Delta\lambda, N)\rangle}{\langle T(\lambda, N)\rangle\langle T(\lambda + \Delta\lambda, N)\rangle} - 1 \qquad (2)$$

where $T(\lambda, N)$ is the intensity of spectral transmission at the $N^{th}$ channels for input wavelength $\lambda$, $\Delta\lambda$ is the spectral spacing between two distinct spots in spectrum, and $\langle ... \rangle$ refers to the average over $\lambda$. The numerical results of $C(\Delta\lambda, N)$ is shown in Fig. **2c**, in which the full-width at half maximum (FWHM) of $C(\Delta\lambda, N)$ reveals the estimated reconstruction resolution. Along with the increase of coupling gaps, the reduction of coupling efficiencies leads to an improvement in $Q_{load}$, thereby narrowing the spectral FWHMs, which leads to the estimated resolution decreases gradually. Spectral responses with sharp and narrow features lead to a higher estimated resolution, therefore, a coupling gap of 180 nm is selected for our chaos-assisted spectrometer.

For comparison of the ray model and real travelling wave properties, Husimi function is applied to the ray dynamics. As a quantum mechanical eigenfunction, Husimi function can convert the wave function in real space into a quasi-probability distribution corresponding to the ray dynamics trajectory state of PSOS in phase space[10]. For a dielectric cavity, the incident Husimi functions[42] are

$$H(s, \sin\kappa) = \frac{nk}{2\pi}\left|Fh_\psi(s, \sin\kappa) \mp \frac{i}{Fk}h_{\partial_\nu\psi}(s, \sin\kappa)\right|^2 \qquad (3)$$

where the weight factor is $F = \sqrt{n}\sqrt{1 - \sin^2\kappa}$ and the overlap functions are

$$h_\psi(s, \sin\kappa) = \int_0^{S_{max}} \xi(s, \sin\kappa, s')\psi(s')ds' \qquad (4)$$

$$h_{\partial_v\psi}(s,\sin\kappa) = \int_0^{S_{\max}} \xi(s,\sin\kappa,s')\partial_v\psi(s')ds' \qquad (5)$$

where $\psi(s)$ and $\partial_v\psi(s)$ is the wave function and its normal derivative, $\xi(s,sin\kappa,s')$ is the minimal-uncertainty wave packet:

$$\xi(s,\sin\kappa,s') = (\pi\sigma)^{-\frac{1}{4}} \sum_{l=-\infty}^{\infty} e^{-\frac{(s'+ls_{\max}-s)^2}{2\sigma}-ink\sin\kappa(s'+ls_{\max})} \qquad (6)$$

where the aspect ratio factor is $\sigma = \sqrt{\sqrt{2}/k}$, controlling the relative uncertainty in $s$ and $\sin\chi$. And $k$ is the wave number.

Fig. **2d** depicts the calculated Husimi maps of four periodic/quasi-periodic resonant modes of the chaotic cavity with $\alpha = 0.375$ by transforming the electric-field distributions into quasi-probability distributions compared with the PSOS. The electric-filed distributions of these mode patterns can be found in Supplementary Information **S5**. The resonances are confined within the stable islands of different period numbers. These high-intensity regions highlighted in four groups of quasi-probability distributions highly dovetail with stability islands in the PSOS, verifying the consistency between the ray model and the travelling wave model. The high-intensity parts in Husimi maps from left to right are separately corresponding to six-bounce orbit with $\sin\chi \sim [0.89, 0.91]$, five-bounce orbit with $\sin\chi \sim [0.85, 0.88]$, five-bounce orbit with $\sin\chi \sim [0.78, 0.83]$, four-bounce orbit with $\sin\chi \sim [0.66, 0.74]$ in ray dynamics.

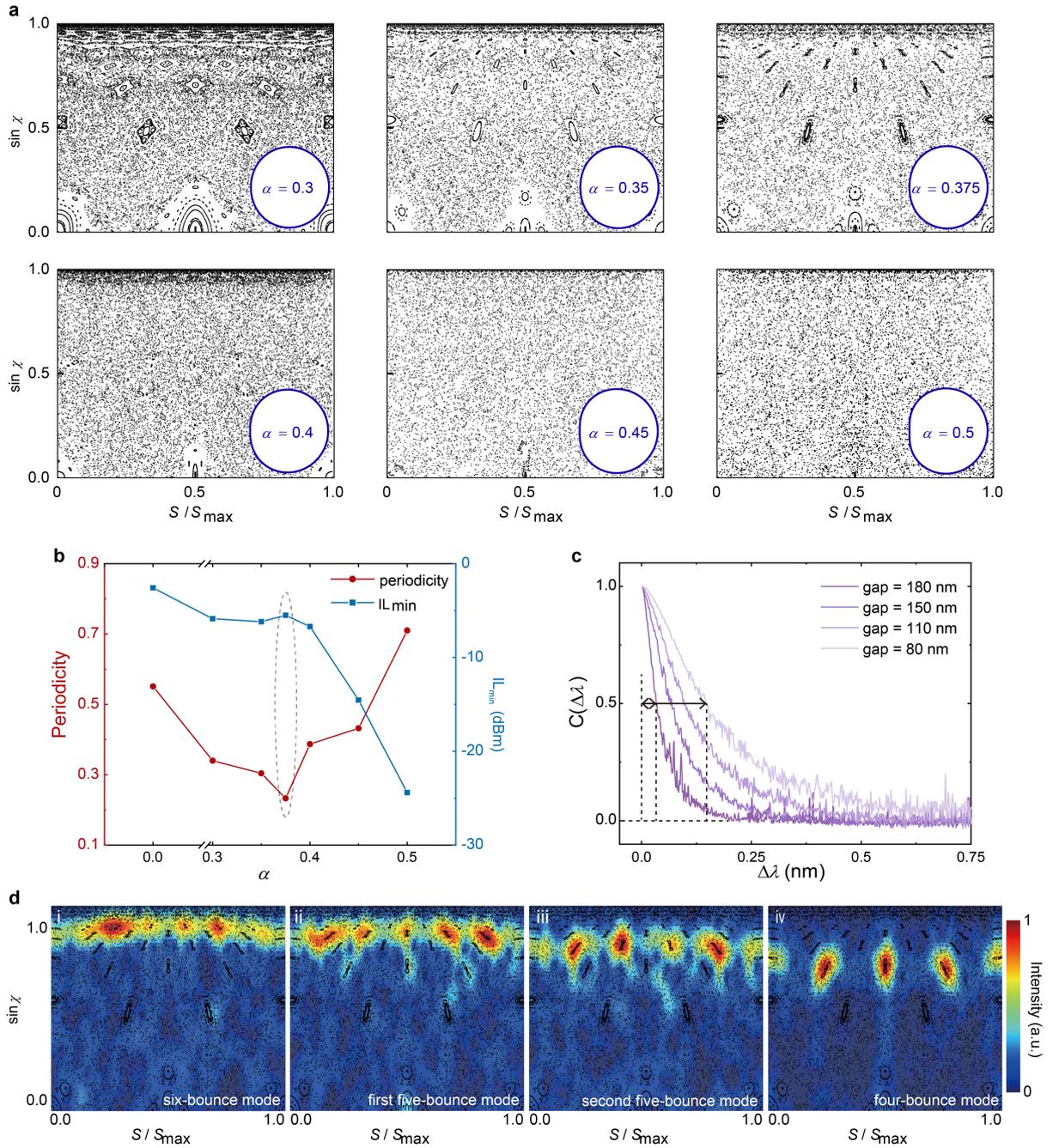

**Fig. 2 | Chaos-assisted spectrometer design. a,** Boundary shape of different deformation parameters $\alpha$ and corresponding calculated PSOS. **b,** Periodicity and insertion loss as functions of a calculated from the measured spectral response of the fabricated devices. **c,** Estimation of the reconstruction resolution, $C(\Delta\lambda)$, around $\Delta\lambda = 0$ for chaotic cavity devices ($\alpha = 0.375$) with different gaps between bus waveguides and chaotic cavity. **d,** Husimi maps of certain periodic/quasi-periodic resonant mode patterns corresponding to PSOS.

**Chaos-assisted spectrometer characterization**

The devices are fabricated with different $\alpha$ and maintaining the same effective radius, in Center for Advanced Electronic Materials and Devices (AEMD) of SJTU (refer to Materials and methods for fabrication details). The scanning electron microscopy (SEM) image of the fabricated chaos-assisted spectrometer of $\alpha = 0.375$ is shown in Fig. **3a**, with the scale bar representing a length of 10 μm. The yellow dashed line signifies a circular shape. The operational region of the chaos-assisted spectrometer occupies only 20×22 μm². The integration of a spiral Titanium (Ti) heater atop the chaotic cavity regions permits the manipulation of all-supported resonant modes of the chaotic cavity through thermo-optic (TO) effects to sample, encode and measure the incident signals. Reconstruction process is illustrated in Fig. **3a.** The optical microscopy image of the fabricated chip is provided in Supplementary Information **S6**. A pre-calibrated self-decorrelated response matrix **T** with $N_p$ rows and $M_w$ columns based on these various spectral features, is obtained by measuring transmission spectra in the output port while linearly sweeping the external heating power applied to the Ti heater. Each row corresponds to a discrete heating power and each column represents a wavelength band. The detected output power **S** with $N_p$ heating channel number is recorded, and the spectral information is sampled and encoded when the unknown incident signal **I** with $M_w$ wavelength points passes through the chaos-assisted spectrometer. **S** can be represented mathematically as a vector product:

$$\mathbf{S}_{N_p \times 1} = \mathbf{T}_{N_p \times M_w} \mathbf{I}_{M_w \times 1} \tag{1}$$

By measuring the optical powers of $N_p$ at values significantly below the Nyquist frequency ($N_p \ll M_w$), it becomes feasible to reconstruct a signal consisting of $M_w$ wavelength points. An estimation of **I** can be obtained by employing a reconstruction algorithm to tackle this inverse problem. The optical microscopy image of the fabricated device is represented in the left part of Fig. **3b**. To mitigate measurement noises caused by unstable electrical contact and spatial oscillation of multi-axis manual stages and adhered optical fibers, we implement optical and electrical packaging for our fabricated chip. The optical microscopy image of the chip after packaging is shown in the right part of Fig. **3b**.

Calibration and characterization of the chaos-assisted spectrometer are conducted. Fig. **3c**

exhibits the normalized response matrix obtained by measuring transmission spectra from the drop port under 300 linearly swept external bias (refer to Supplementary Information **S6** for the transmission spectrum of the reference waveguide). The normalized transmission spectra under the 1$^{st}$ and 300$^{th}$ heating channels are plotted in Fig. **3d** (refer to Materials and methods for calibration process). The loaded Q-factors of certain resonances, indicated by black arrows in Fig. 3**b**, are summarized in Supplementary Information **S7**, as well as the discussion of Q-factor. These narrow resonance peaks with high Q-factors, exhibit sharp transmissions with significant variations at different wavelength points near resonance peaks, forming the foundation of achieving high reconstruction resolution. The average auto-correlation is calculated and plotted in Fig. **3e**. Measured as 0.210, the effectively suppressed periodicity indicates sufficient diversity has been introduced into the response matrix through chaotic effect. The cross-correlation function is employed to assess the degree of mutual independence of each transmission spectrum in the response matrix. The red line in Fig. **3f** represents the computed average cross-correlation which consistently maintains at a low level below 0.08, suggesting an almost uncorrelated response matrix with good orthogonality. Three other cross-correlations for arbitrarily selected pairings of transmission spectra are also included in Fig. **3f**, marked as green, blue, and yellow scatter dots. The temporal sampling decorrelation is further validated, detailed in Supplementary Information **S8**. The estimated resolution calculated according to equation (2) is characterized in Fig. **3g**, measuring approximately 40 pm. The estimated resolution roughly conforms to the average level of Q-factors throughout the transmission spectrum (estimated resolution for microdisk resonators can be found in Supplementary Information **S9**).

Furthermore, we perform numerical analysis on the pre-calibrated response matrix **T**, detailed in Supplementary Information **S10**. The calculated condition number and singular values validate the good orthogonality of our chaos-assisted spectrometer. Fast Fourier transform of left singular vector enabling the potential collection of comprehensive information with fewer channels. Discrete *Picard* condition is satisfied for several probe signals, conforming the existence of bounded and convergent solution with practical norms. Numerical computations are also conducted to evaluate the effectiveness of the chaos-assisted spectrometer in extracting different types of signals, including discrete, smooth, and step signals (refer to Supplementary Information **S11** for the test signals and reconstructed signals). Noise tolerance of the response matrices is also analyzed, refer to Supplementary Information **S12**. In addition, we evaluate the reconstruction quality and resolvability of our chaos-assisted spectrometer with the optimal deformation parameter $\alpha = 0.375$ in comparison to other devices, including microring and microdisk resonators, as well as chaotic cavities with $\alpha = 0.3$ and $0.45$. We examine the quality of these response matrices by a discrete signal with adjacent multi-peak signals in the wavelength domain, as well as a sparse discrete signal with varying peak intensities across the whole operational bandwidth, with different noise levels loaded into the measurement. The spectrometer based on the chaotic cavity with $\alpha = 0.375$ delivers the most superior reconstruction performance and strongest robustness against measurement noise compared to other conventional resonators and chaotic cavities with different $\alpha$, coinciding with its smallest condition number. The superiority of the chaotic cavity over conventional resonators (microring and microdisk resonators) and the optimal deformation parameter of 0.375 are further validated. See details in Supplementary Information **S13**.

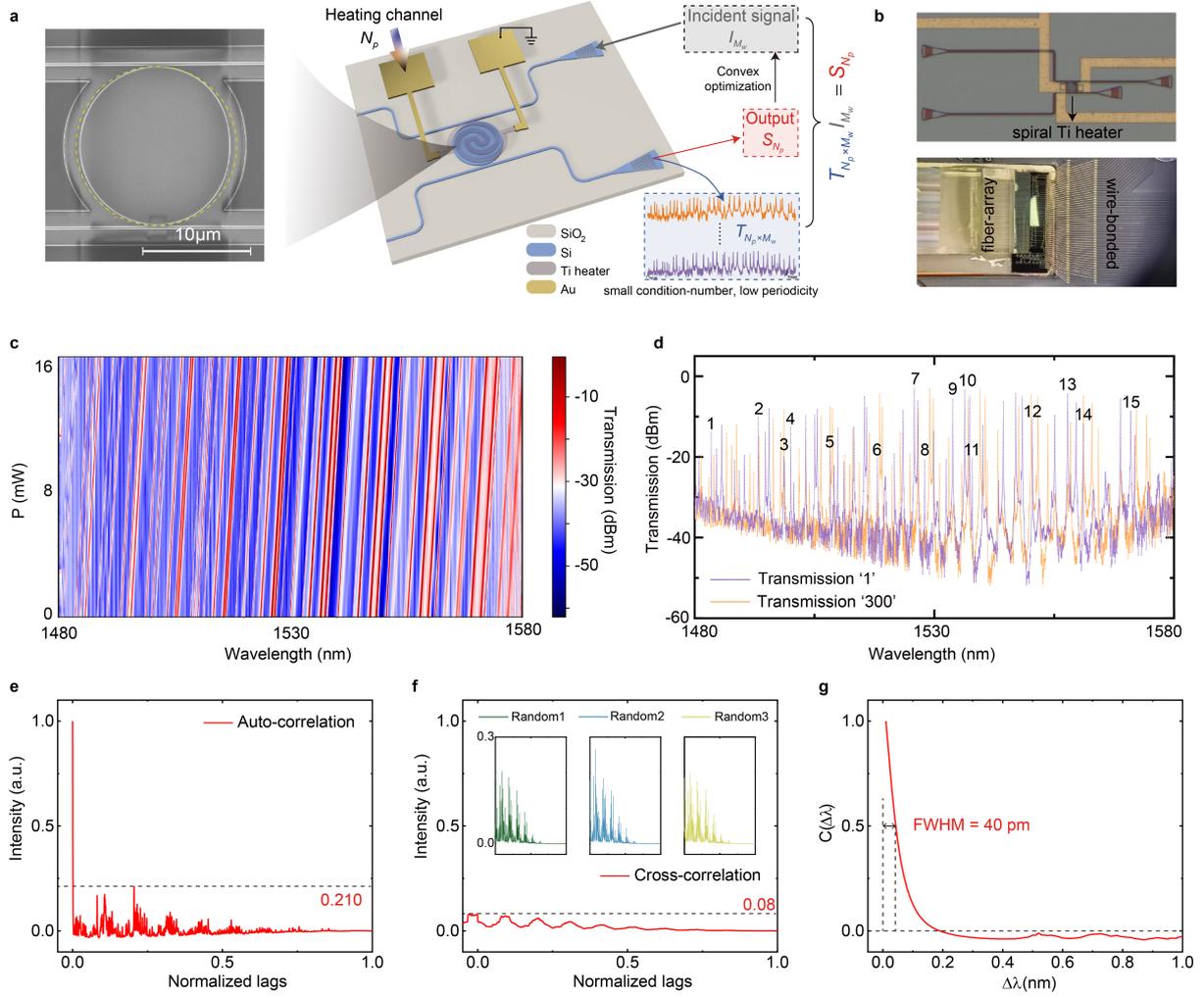

**Fig. 3 | Calibration and characterization.** **a,** Schematic of chaos-assisted spectrometer and the spectral reconstruction process. The zoom-in figure shows SEM photo of the functional region of the fabricated chaos-assisted spectrometer, while the scale bar indicates a length of 10 μm. The yellow dashed line indicates circular shape. **b,** Upper: optical microscopy photo of the fabricated device after Ti heaters and Au interconnection layer deposition. Bottom: Image of the photonic chip wire bonded to a customized PCB board with a fixed fiber-array. **c,** Normalized response matrix obtained by sweeping the heating power linearly from 0 to about 16.5 mW from 1480 nm to 1580 nm with 300 heating channels. **d,** Normalized transmission spectra at 1st and 300th heating channels. Black arrows and numbers mark resonance peaks detailed in Supplementary Information Table. **S1**. **e,** Auto-correlation function where the periodicity level is shown by the black dashed line. **f,** Cross-correlation function. The red line is the average cross-correlation function. Inserted graphs are the cross-correlation functions of three randomly selected transmission groups, designated as green, blue, and yellow. **g,** Spectral

correlation function of $C(\Delta\lambda)$ around $\Delta\lambda = 0$ where the estimated resolution is illustrated by the black dashed line.

**Experimental results**

We then conduct experimental validation of chaos-assisted spectrometer performance with respect to incident signals with different optical features. We first launch narrowband signals from a tunable laser source whose wavelengths are set throughout the operation bandwidth from 1480 nm to 1580 nm for single-peak reconstruction. The heater is driven by the varying voltages of an automatically programmed electrical source and the sampled optical powers associated with heating channels are measured simultaneously. The input narrowband signals can be regarded as discrete signals marked as dashed black lines owing to the 60 kHz ultra-narrow bandwidth. Relative errors ($\varepsilon = \frac{\|\mathbf{I}^\dagger - \mathbf{I}\|_2}{\|\mathbf{I}\|_2}$) and coefficients of determination ($r^2 = 1 - \frac{\sum_1^n (I_i^\dagger - I_i)^2}{\sum_1^n (I_i - \bar{I})^2}$) are employed to assess the accuracy of the reconstructed spectra. The reconstructed results of single narrow-band signals are depicted in Fig. **4a** and zoom-in spectra are presented in the second row. The signals within the operational bandwidth are accurately reconstructed with precise positioning and consistently maintain a FWHM of approximately 10 pm. Furthermore, the operational bandwidth of the spectrometer is constrained by the physical properties of the material systems and limited by the experimental spectral response of the grating couplers (GC) and the measuring instruments.

By simultaneously launching two laser signals at different wavelengths, we further investigate the resolution limits. Fig. **4b** illustrates the reconstruction result at wavelength intervals of two laser sources of 10 pm, 20 pm, and 100 pm, respectively, as blue dashed lines. The peak intensity interval of 10 pm gap is clearly distinguishable with a low relative error of 0.078, demonstrating the ultra-high reconstruction resolution of 10 pm in our chaos-assisted spectrometer. As summarized in Table. **S1**, the Q-factors are independent of the wavelength, exhibiting a basically consistent performance across our entire operational bandwidth, stating consistent spectral perturbations across the whole operational bandwidth. We further conduct more experiments for dual-peak signal reconstruction at various wavelength positions covering operational bandwidth to validate the consistency of ultra-high resolution for our chaos-assisted spectrometer throughout the whole operational bandwidth. These dual-peak discrete signals yield a wavelength separation of 10 pm. As illustrated in Fig. **4c**, all the reconstruction results for 10 pm interval double-peak signal at 1493.1 nm, 1508.5 nm, 1548.8 nm, and 1562.9 nm exhibit high accuracy and most reconstruction errors are lower than 0.1. Clearly, it's evidenced that our chaos-assisted spectrometer can support the stable resolution ability of 10 pm at the whole operational bandwidth.

We continue to address the reconstruction of a multi-peak signal that span the whole operational bandwidth, featuring peaks situated at the central and marginal regions of the operational wavelength range, thereby verifying the consistent performance of our chaos-assisted spectrometer, as shown in Fig. **4d**. The reconstruction results could reveal a low relative error of 0.122, a high coefficient of determination of 0.982, and a peak signal-to-noise ratio (PSNR) of approximately 14 dB, fully testifying the reconstruction performance of our chaos-assisted spectrometer across the whole operational bandwidth.

For better verifying the comprehensive spectra reconstruction performance of our chaos-assisted spectrometer, we further supply more experimental reconstruction results for various complex spectrum patterns. A waveshaper is utilized to encode a continuous signal from an Erbium Doped Fiber Amplifier (EDFA), and the spectrum is pre-measured by a commercially available optical spectrum analyzer (OSA, Yokogawa AQ6370C) for reference (refer to Materials and methods for detailed measurement). Reconstruction result for this continuous signal, characterized by Sinc function waveform attributes, is illustrated in the left panel of Fig. **4e**, with a calculated $r^2$ of 0.986 and $\varepsilon$ of 0.099. A smooth continuous signal directly emitted from an EDFA over a wider bandwidth of 90 nm is reconstructed, as depicted in the right panel of Fig. **4e**, with a calculated $r^2$ of 0.980 and $\varepsilon$ of 0.120. These resolved spectra and reference spectra measured by OSA are represented by red solid line and black dashed line, respectively. More reconstruction results for a continuous bandpass signal, as well as for hybrid spectra are provided in Supplementary Information **S14**. The primary source of disagreement in continuous signal reconstruction stems from the fact that continuous signals do not conform to the sparse representation required by compressed sensing theory, particularly under our high compression ratio conditions, which is the inherent reason why solving for continuous signals is significantly more challenging than for discrete signals. Besides, discretization error becomes more pronounced in our chaos-assisted spectrometer with widely distributed high-Q resonant peaks. While this predicament can be mitigated by increasing the number of sampling wavelength points using a finer grid, it would concurrently magnify the demand of sampling channels and the computational burden to a large extent. Additionally, the optical and electrical noises from the mechanical vibration of measuring devices and electrical contacts, as well as the temperature fluctuations of the measurement process, are significant causes of reconstruction errors. Temperature controllers will be considered to mitigate temperature fluctuations in our future research. Detailed theoretical analysis of reconstruction error for the continuous signal can be found in Supplementary Information **S15**.

Additionally, we further investigate more characteristics of spectral reconstruction. Dynamic range is validated through the high-accuracy reconstruction of a dual-peak discrete signal with significant intensity disparities, encompassing one peak in the central wavelength region and another in the longer wavelength region (see Supplementary Information **S16**). Stability considering electrical fluctuations, temperature effects, as well as overall experimental reconstruction test and analysis for stability are detailed in Supplementary Information **S17**. The electrical current of our chaos-assisted spectrometer exposes only 0.1% fluctuation after 7 days of the initial measurement, and the spectrometer would experience a redshift of 0.4 nm under a temperature rising of 5 K, indicating that the incident spectrum could still be correctly reconstructed with approximately $\pm 0.25℃$ temperature variation.

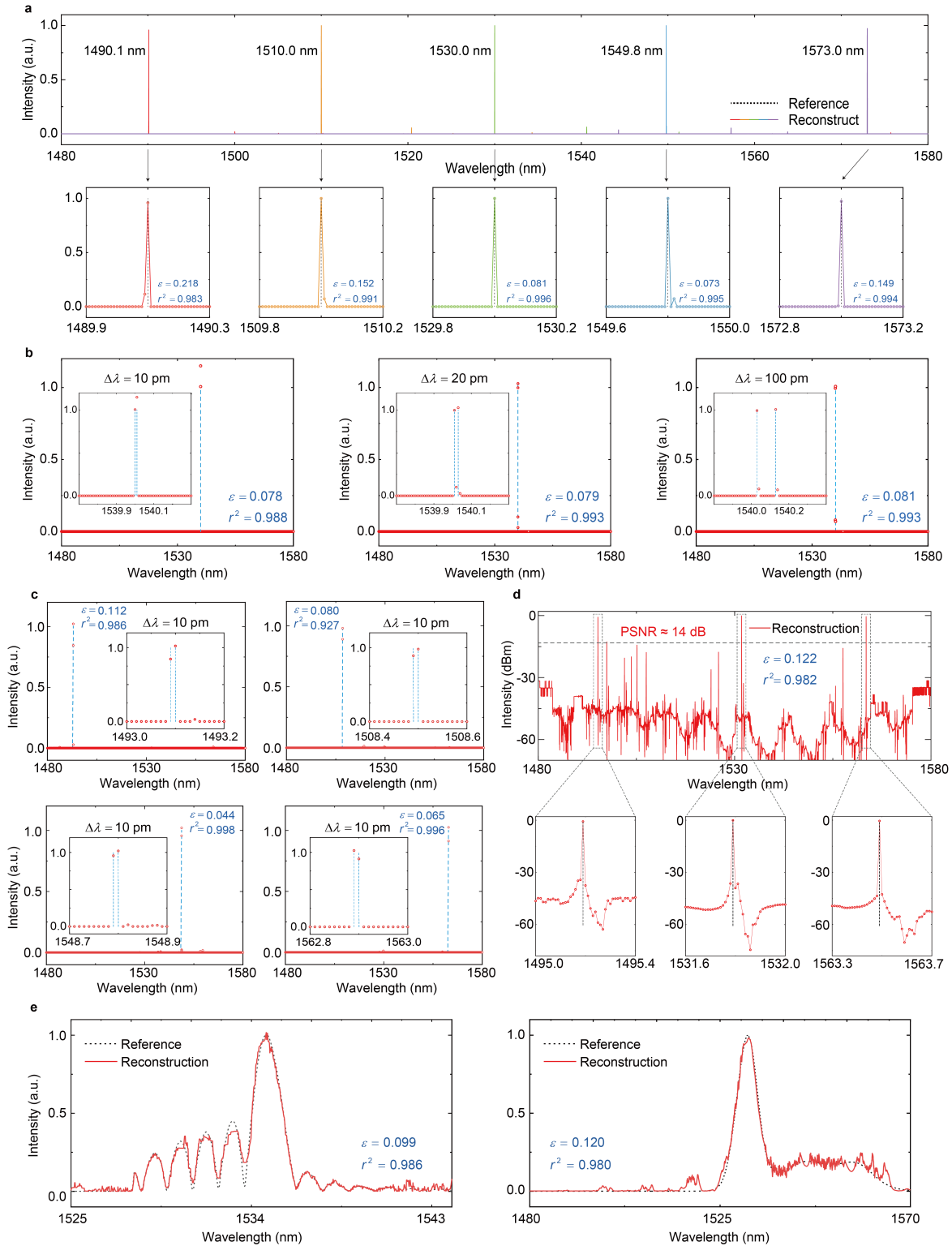

**Fig. 4 | Reconstruction results. a,** Narrow, monochromatic peak signals that each across the entire operational range from 1480 nm to 1580 nm; **b,** Double-peak laser signal with different wavelength intervals of 10 pm (left), 20 pm (middle), and 100 pm (right), validating resolution of 10 pm, with each signal spanning the entire operational bandwidth. **c,** 10 pm interval double-

peak laser signals at (top left) 1493.1 nm, (top right) 1508.5 nm, (bottom left) 1548.8 nm and (bottom right) 1562.9 nm, with each signal spanning the entire operational bandwidth. **d,** A multi-peak signal with peaks situated at the central and marginal regions of the entire operational wavelength range. **e,** Continuous signal with Sinc function waveform (left), and broad EDFA signal (right).

**Discussion**

Shrinking the area while maintaining the high performance will be a key focus for the next generation of miniaturized computational spectrometers. These devices, used in potential applications like non-invasive medical diagnostics in wearable health monitors, or on-the-go integrated spectrometer in smartphones, will revolutionize food safety testing, counterfeit detection, and personal health monitoring, etc. Concurrently, as various on-chip muti-spectral imaging chips which generally rely on the optical components capable of processing spectral information are recently reported[43–46], a potential brand-new demand for high-performance integrated spectrometers are emerging. In reality, the biggest spectrometer miniaturization limitations rarely come from electronics, as the transistors can be made down to scale of nanometers, while the on-chip spectrometers with adequate performance are mostly measured in hundreds of microns or even millimeters in critical dimension[26,33,34,47,48]. Shrinking these "blank spaces" between electronics (for spectra reconstruction) and photonics (for spectra encoding) requires genuine physical innovations in spectra manipulation, and will be at the heart of making spectrometers smaller, cheaper, and higher density of functionalities[49].

Exploiting the chaotic behavior of the spectrum induced solely by smooth microcavity deformation, our approach yields a complex spectral response that effectively suppresses periodicity found in conventional resonators, thereby introducing diversity and generating highly de-correlated response matrix in a simple way that are long sought for state-of-art computational spectrometers in various complicated configurations. We have experimentally demonstrated a chaotic computational spectrometer boasting ultra-high resolution (10 pm) and a wide operational bandwidth (100 nm) within an incredibly compact footprint of merely 20×22 μm². A comprehensive performance evaluation of various state-of-art computational spectrometers which have been previously demonstrated is executed in Fig. **5** (Another form of the comprehensive performance evaluation refers to Supplementary Information Fig. **S29**). Owing to the buildup of differences in the optical path length, the significant effect of footprint on resolution is necessary to be considered in addition to the trade-off between resolution and bandwidth. We emphasize the superior performance of our chaos-assisted spectrometer in terms of bandwidth-to-resolution ratio per unit footprint, defined as BRFR (bandwidth/resolution/footprint). This ratio is one of the most representative indicators of spectrometer performance, regarding the intrinsic contradiction between the higher OPL for high resolution and minimizing footprint while maintaining large operational bandwidth. Up-to-date, we have most thoroughly overcome this typical three-way trade-off and realize a record-high BRFR of 22.7 μm$^{-2}$, surpassing other computational spectrometers by one to four orders of magnitude (refer to Supplementary Information **S18** for details).

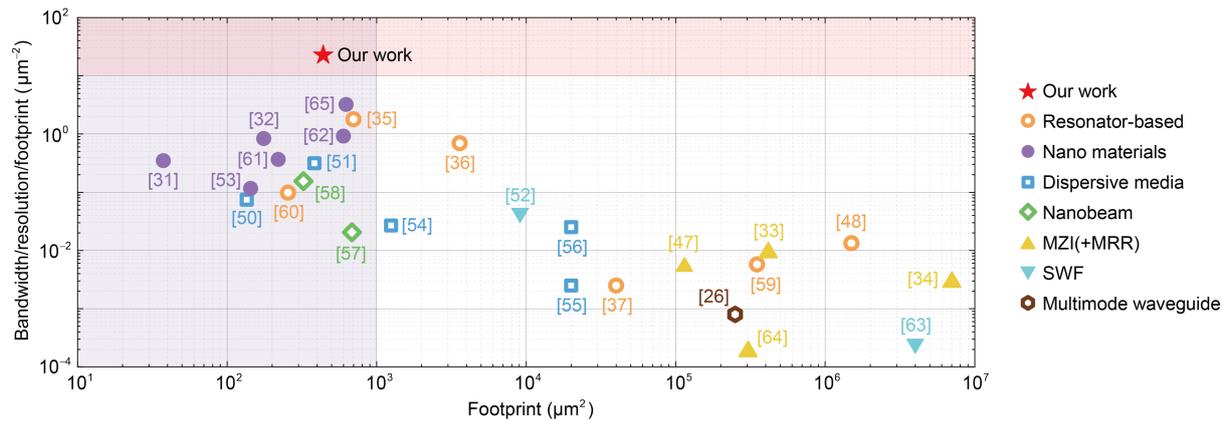

**Fig. 5 | Performance comparison.** Comparative evaluation of the performance between our work and other prominent computational spectrometers, where MZI, MRR and SWF are abbreviations for Mach-Zehnder interferometer, microring-resonator and stratified waveguide filters, respectively[26,31,33–37,47,48,50–65].

Furthermore, the high compression ratio of our chaos-assisted spectrometer with a value of 33.33 is emphasized, surpassing most previous works (See details in Supplementary Information **S19**). In addition, the chaos-assisted spectrometer only uses a single spatial channel, further compacting the physical size. Despite GC being prevalent in chip testing phases, limited by the narrow bandwidth and high loss of GC due to the inherent diffractive characteristics, the operational bandwidth of our chaos-assisted spectrometer is curtailed to about 100 nm (see spectral response of GC in Supplementary Information **S6**). The substitution of GCs with edge couplers permits a substantial expansion of the operational bandwidth to over 300 nm, with insertion losses of edge couplers being less than 1.86/2.80 dB in O/S+C+L band[66]. This proposal can be transplanted into other wavelength bands via simply adjusting the dimension of the chaotic cavity in accordance with wavelength among silicon transparent window, or transplanted into other material systems, as detailed in Supplementary Information **S20**. Unlike conventional computational spectrometers relying on topological arrangement of interferometers, dispersive optics or resonators to generate specific spectra, our disruptive chaotic computational spectrometer operates independently of intricate optical components or complex system setups. Moreover, current on-chip computational spectrometers generally cost high power consumption of over 30 mW, as summarized in Table. **S5**. Comparatively, our chaos-assisted spectrometer only requires a low power consumption of 16.5 mW. Thus, our work represents a major step towards the extreme miniaturization of spectrometers and potentially facilitates on-chip, low power consumption, cost-effective spectral imaging.

For the computing resources, the optimization for the linear inverse problem process can be expedited on a moderately performing personal computer (close to some common mobile devices) with several seconds. Factually, for the system stability of our spectrometer, we have to cost a lengthy time-cost to finish the whole process including measurement, communication and reconstruction (details refer to **Materials and Methods**), lacking the directly potential for the real-time applications. Nevertheless, the electro-optic modulation mechanism employment with comprising over 1000 points within $1 \times 10^{-7}$ seconds would substantially alleviate time consumption in constructing the response matrix, and the more high-speed electrical interconnections by the more advanced controller modules could effectively mitigate the limited data transmission rate. Meanwhile, the proper upgrade from our used medium or low-end computer hardware (details refer to **Materials and Methods**) could assist to further reduce the time-cost in the post processing for the reconstruction process in our spectrometer. Additionally, there exists a limitation to further miniaturization. The ray dynamics model is tenable when the cavity scale exceeds the wavelength of the incident light. Theoretical analysis based on the ray dynamics model becomes unreasonable when the size of chaotic cavity continues to shrink. Meanwhile, a more compact cavity results in the decline of the resonant modes to hinder the spectrometer performance. Detailed performance comparison of chaotic cavities with effective radii of 5 μm and 10 μm can be found in Supplementary Information **S21**. Thus, the chaotic cavity with 10 μm efficient radius currently employed for our chaos-assisted spectrometer, signifies the limit for miniaturization.

In addition, the chaotic cavities demonstrate good reproducibility. The consistency of deformation profiles guarantees the consistency of profiles of spectral responses, thus ensuring identical spectrometric performance metrics across various devices. Yet, minor wavelength shifts are observed, which can be attributed to nonuniformity in the fabrication process. During the potential commercial deployment, the occurrence of slight wavelength shifts can be uniformly corrected by heating the Ti heater with an external power through TO effect, aligning the resonant wavelengths of various devices integrated on the chip. Details are provided in Supplementary Information **S22**.

## Materials and Methods

### Chip Fabrication

The devices are fabricated on a silicon-on-insulator (SOI) wafer with a 220 nm top Si layer on a 3 μm buried $SiO_2$ layer. The Si waveguides and gratings are firstly patterned by the electron beam lithography (EBL) system and then fully etched by using a single-step inductively coupled plasma dry etching. 1 μm top $SiO_2$ passive layer is deposited subsequently by plasma-enhanced chemical vapor deposition. Ti metal heater with a thickness of 120 nm and Au metal interconnection with a thickness of 350 nm are then defined by EBL and deposited using the electron beam evaporator in sequence.

### Calibration and reconstruction experiments

A pre-calibration process is demonstrated as follows. A tunable continuous wave laser (Santec TSL 770) and a power monitor (Santec MPM 210) are utilized for sampling and data collection. Broadband GC with less than 5 dB insertion loss are applied for fiber-chip coupling. A source meter (Keithley 2400) is used to offer an external driving power source. A maximum external power $P_{max}$ of about 16.5 mW is applied to the heater. For eliminating current fluctuations due to unstable contact, we utilize wire bonding, an electrical package and optical package for our fabricated chip. The current value can be stabilized at $1\times10^{-4}$ mA after electrical packaging. The dimensions of response function **T** are determined by the wavelength point numbers $M_w = BW/\delta\lambda$, and $N_p = P_{max}/\delta P$ mutually, where $N_p$ is the sampling channels of 300; $BW$ is the spectrometer measurable bandwidth; $\delta\lambda$ is the wavelength grid of 10 pm. Here, a spectrum with $M_w$ of 10000 points can be reconstructed with merely $N_p$ of 300 heating power channels, demonstrating an ultra-high compress ratio.

For the single peak reconstruction, we use a CW laser source (Santec TSL 770) to generate discrete narrow linewidth signals with wavelengths of 1490.1 nm, 1510.0 nm, 1530.0 nm, 1549.8 nm, and 1573.0 nm and inject them into the input port of the chaos assisted spectrometer, demonstrating an operation bandwidth of 100 nm. For two closely separated discrete signals reconstruction, two narrow linewidth signals with wavelength separation of 10 pm, 20 pm, and 100 pm are produced by two CW lasers which combined with a 3-dB coupler subsequently. The wavelengths of two CW lasers are set as: 1540 nm and 1540.01 nm (10 pm), 1540.01 nm and 1540.03 nm (20 pm), 1540.03 and 1540.13 nm (100 pm), respectively.

For continuous signal reconstruction of Sinc function waveform, A programmable optical filter (Finisar Wave-shaper 1000s) is utilized to encode the continuous signal emitted from an Erbium-Doped Fiber Amplifier (EDFA) to generate the required test signals. A polarization controller is connected subsequently to adjust the continuous signal to TE mode, and a 3-dB coupler divides the signal into two paths: one is injected into the chaos-assisted spectrometer while a commercial optical spectrum analyzer (OSA, Yokogawa AQ6370C) records spectra from the other path. For the broadband continuous signal reconstruction, an EDFA from 1530 to 1565 nm is directly used to provide a broadband continuous signal. The number of temporal sampling channels is increased to 1000. The spectrum reconstructions are performed by running the CVX optimization algorithm on MATLAB by the "Mosek" solver, based on an AMD Ryzen 7 3700X CPU and NVIDIA GeForce GTX 1650 with 32 GB memory. The reconstruction of continuous signals is achieved in approximately 3 seconds, whereas the reconstruction of discrete sparse signals requires approximately 2 seconds. The pre-calibration process takes approximately 750 s. For the spectrum measurement process, we sample every 0.5 s with an integration time of 0.1 s. Since the rise time and the fall time are less than 10 μs, sampling starts at 0.3 s after sending commands to the source meter to regulate the external biases, in order to ensure a stable temperature of silicon waveguide. ANSYS Lumerical FDTD is used to perform the optical transmission simulations for proposed the chaos assisted spectrometer.

**Reconstruction algorithms.**

The reconstructed signal $\mathbf{I}^\dagger$ can be generated by solving the under-determined linear least-squares:

$$\mathbf{I}^\dagger = \mathrm{argmin}_{\mathbf{I}\in\mathbb{R}^+}\|\mathbf{TI} - \mathbf{S}\|_2^2 \tag{14}$$

where $\|\cdot\|_2$ represents the $l_2$-norm. It is noteworthy to mention that the response matrix obtained in real-world circumstances will surely encounter measurement noise plus a certain level of ill-conditioning because of the fewer sampling points. In order to overcome this overfitting obstacle, the regularization coefficient is introduced as:

$$\mathbf{I}^\dagger = \mathrm{argmin}_{\mathbf{I}\in\mathbb{R}^+}(\|\mathbf{TI} - \mathbf{S}\|_2^2 + \alpha_1\|\mathbf{I}\|_1 + \alpha_2\|D_1\mathbf{I}\|_1 + \alpha_3\|D_2\mathbf{I}\|_2), 0 \leq \mathbf{I} \leq 1$$

where $a_1$ is the weight regularization coefficient for the $l_1$-norm of input matrix $\mathbf{I}$, which is vital in the regression of $\mathbf{I}^\dagger$ into discrete untrivial solutions; $D_1$ and $D_2$ refer to the first and second derivative operator; $a_2$ is the weight regularization coefficient for the $l_1$-norm of $D_1\mathbf{I}$; $a_3$ is the weight regularization coefficient for the $l_2$-norm of of $D_2\mathbf{I}$, which is critical for optimizing continuity and smoothness of continuous broad-band signals.

**Acknowledgments**

This work was financially supported by the National Key R&D Program of China (2023YFB2804702); Natural Science Foundation of China (NSFC) (62175151, 62341508); Shanghai Municipal Science and Technology Major Project. We also thank the Center for Advanced Electronic Materials and Devices (AEMD) of Shanghai Jiao Tong University (SJTU) and United Microelectronics Center (CUMEC) for fabrication support.

**Conflict of interests**

The authors declare no competing interests.

**Author contributions**

X.H.G initiated the project. C.J.X, Y.J Z performed the calculation and simulation. Y.J. Z. and X.H.G. designed the experiments. Y.J Z. fabricated samples. Y.J Z. and Z.Y. Z. carried out the measurements. Y.J Z, C.J.X, Z.Y.Z, Y.K.S and X.H.G analyzed the results and wrote the manuscript. X.H.G and Y.K.S. supervised the project.

**Correspondence and requests for materials** should be addressed to Xuhan Guo.

**Data availability**

The data that support the plots within this paper are available from the corresponding authors upon request.